# Concave risk measures in international capital regulation

**Imre Kondor[1], András Szepessy[2], and Tünde Ujvárosi[2]**

[1]Department of Physics of Complex Systems, Eötvös University, 1117 Budapest, Pázmány P. sétány 1A, Hungary, and Collegium Budapest – Institute for Advanced Study, 1014 Budapest, Szentháromság u. 2,
[2]Raiffeisen Bank, 1054 Budapest, Akadémia u. 6, Hungary

**Abstract**

We show that some specific market risk measures implied by current international capital regulation (the Basel Accords and the Capital Adequacy Directive of the European Union) violate the obvious requirement of convexity in some regions in the space of portfolio weights.

## 1. Introduction

Any reasonable risk measure must be convex. A non-convex risk measure penalises diversification, does not allow risk to be correctly aggregated, cannot provide a basis for a rational pricing of risk, and cannot serve as a basis for a consistent limit system. In fact, a non-convex risk measure is not a risk measure at all.

The concept of risk measures has undergone considerable evolution over the past couple of years. In a Gaussian world, or, slightly more generally, in an elliptical world, where the level surfaces of the joint probability distributions of financial assets are hyperellipsoids (Cambanis et al., 1981), standard deviation is essentially the unique measure of risk. It is a positive definite quadratic form in the portfolio weights, hence its level surfaces are hyperellipsoids (and, as such, convex) themselves. Dramatic events, like the 1987 Black Monday, have convinced the financial industry, however, that normal statistics can be very far from market reality. With this, variance fell out of favour, and came to be replaced by value at risk (VaR) (JP Morgan 1994, Jorion 2000), quickly embraced by industry and regulation alike. VaR has serious shortcomings of its own, however: as a quantile, it has no reason to be convex, and indeed, it behaves quite erratically upon aggregation (see e.g. Danielsson et al 2001, or the papers by Acerbi and Tasche 2002; Frey and McNeil 2002; Rockefellar and Uryasev 2002, and others in the special issue "Beyond VaR" of Banking & Finance, 26, July 2002, for a critique of VaR). As a remedy, a simple set of axioms any coherent risk measure should satisfy was introduced a few years ago (Artzner et al, 1997, 1999). Independently and almost simultaneously similar ideas were put forward by a number of workers (Wang et al, 1997; Hodges, 1998; Carr et al, 2001). The idea of coherent measures has gained widespread acceptance in academia; some of its simple implementations (e.g. CVaR) have started to appear even in advanced risk monitoring methodologies.

Risk measures are a major concern also for capital regulation. Own capital is generally regarded as a cushion against risk. The minimal capital requirements are now fixed by internationally accepted rules, dating back to the first Basel Accord of 1988. Its European implementation is the Capital Adequacy Directive of the ECC (CAD 1993), issued in 1993 and amended in 1998 (CAD 1998). It has since become law in the member states, and in the accession states as well. Basel I is under revision and will be replaced by a new Accord that is the subject of intensive debate across the industry (see the home page of the Basel Committee at www.bis.org). While the new regulation will introduce substantial changes regarding e.g. credit risk, its rules pertaining to *market risk* are to stay basically unmodified, and these are the ones we will be concerned with here.

Originally, CAD defined a set of detailed algorithms to calculate the capital charges associated with the many different kinds of positions in the trading book. This set of algorithms constitutes what is called the *standard model* now. The 1998 Amendment was introduced with the purpose to allow banks to develop their own internal models and calculate the minimal required capital from them. However, the regulation stipulates stringent rules concerning these own models, and a factor 3 to 4 to multiply the capital requirement resulting from them. In view of this, several European banks have decided to keep to the standard model for reporting purposes, and set up an internal model of varying sophistication for the sake of their own risk management.

The capital charges assigned to various positions in the standard model purport to cover the risk in those positions. Therefore, they must be regarded as some kind of risk measures. Our purpose here is to take a close look at these regulatory risk measures. We are not going to discuss whether they capture the given kind of market risk in an adequate manner, instead we analyse them exclusively from the point of view of convexity. As we shall see, the level surfaces of constant regulatory capital, or, by implication, of constant risk are given by some hyperpolyhedra in the space of portfolio weigths, and they are obviously meant to represent crude approximations to the "true" risk surfaces. If chosen properly, these polyhedra should be convex. We explicitly construct these polyhedra below, and check them for convexity.

Our findings are mixed: The capital requirements of the specific risk of bonds and also those of the FX portfolio define two different regulatory polyhedra which are convex, as they should be. Depending on the composition, however, the level surfaces of risk for an equity portfolio can become concave and, paradoxically, this is due precisely to the special rules which were designed to forbid (as in the original CAD) or penalise (as in some of its national implementations) the excessive concentration of equity portfolios.
Likewise, the regulatory surface for the general risk of bonds can be deeply concave. As a result, one can easily construct model portfolios for which a smaller exposure attracts much larger capital than a larger one. In addition, for some portfolios the capital requirement exhibits wild fluctuations due to the transition of the portfolio components between different maturity zones. The violation of convexity and the instability of the risk measure for the bond portfolio are due to the improper choice of some risk weigthings in the regulation. We map out the space of these weights and identify the subspace where convexity holds, i.e. where consistency could be restored by a minimal modification.

## 2. Risk measures implied by the trading book regulation

We are going to discuss the capital charges associated with four different types of instruments as described in CAD: with the specific and general risk of bonds, with the foreign exchange portfolio, and with the equity portfolio, respectively. As a legal document, CAD is mostly formulated in words. The first task is to construct the mathematical formulae corresponding to the verbal rules that define the capital charges. Space limitations do not permit us to quote the complete wording of the various rules we translate into formulae, and the reader is referred to the original text (CAD 1993) if she wishes to check the correspondence between the legal and mathematical formulations.

2.1. Specific risk of bonds

The corresponding rule is described in Annex I, §14, of CAD. This rule tries to capture the risk due to issuers and to maturities of the bonds in the trading portfolio.
The capital charge is calculated from the formula

$$A = \sum_{i=1}^{n} \gamma_i |x_i| \qquad (1)$$

where $\gamma_i$ is some weight depending on the issuer and on maturity. (The weight is 0,00% for central government items; 0,25%, 1,00%, and 1,60% for qualifying items with residual maturities up to 6 months, between 6 and 24 months, and over 24 months, respectively; and 8,00% for other items.) The $x_i$ are positive (long) or negative (short) net positions in item $i$.
The "iso-risk" surface $A$ = const. is a closed polyhedron (in 3 dimensions an octahedron, shown in Fig.1) which can easily be seen to be convex for any $\gamma_i \geq 0$ and any number of dimensions (number of different items) $n$.

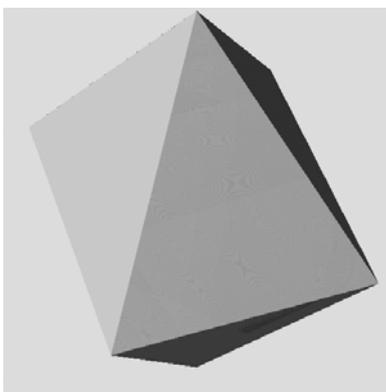

Fig.1: The level surface of the capital requirement associated with the specific risk of a portfolio consisting of three bonds. The weights have been absorbed into the positions. Note that here, and also in the subsequent figures, the risk polyhedra are shown so as to make clear their convexity properties, disregarding their actual position in the coordinate system.

2.2. Foreign exchange

The rules pertaining to foreign exchange risk are described in Annex III of CAD. According to §1, if an institution's overall net foreign-exchange position exceeds 2% of its total own funds $K$, then the capital requirement against FX risk will be 8% of the excess. The definition of the overall net foreign-exchange position $Z$ is given in §4: it is the larger of the total of the net short positions

$$S = \sum_i \frac{1}{2}(sign(x_i)-1)x_i \quad , \quad (S \geq 0)$$

and the total net long positions

$$L = \sum_i \frac{1}{2}(sign(x_i)+1)x_i \quad ,$$

i.e. $Z = \max\{L, S\}$.

Now $L$ and $S$ are obviously related to the gross position

$$G = \sum_i |x_i|$$

and to the net position

$$N = \left|\sum_i x_i\right|$$

by $G = L + S$, and $N = |L - S|$, which gives

$$Z = \frac{1}{2}(G+N)$$

Accordingly, the capital requirement can be calculated as

$$F = 0.08 \max\left\{\left(\frac{1}{2}(G+N) - 0.02K\right), 0\right\} \tag{2}$$

When $F$ is not zero, its iso-risk surface is given by $G + N = $ const. In 3 dimensions the corresponding polyhedron is shown in Fig.2. Again, it can be shown that the surface $F$ = const. is convex in any dimensions.

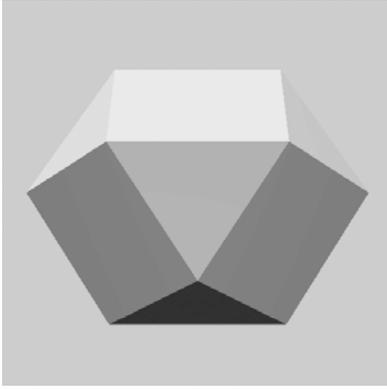

Fig.2: The iso-risk surface of a three component FX portfolio.

### 2.3. Equity risk

According to Annex I, §32 and §34, the capital requirement of the specific and general risk of equities is proportional to yet another combination of $G$ and $N$, namely to $C = \tfrac{1}{2}G + N$. In the original CAD the capital charge was defined as 8% of $C$. Under certain conditions, described in §33, the competent authorities may allow the factor in front of $G$ to be reduced to ¼, but we cannot go into these fine details in here. More important for our present purposes is the fact that, at variance with the original directive, in some of the national implementations of CAD (including the British or the Hungarian ones, e.g.) large positions (exceeding 20% of the total) add an extra term to the formula for the capital requirement. The complete expression for the capital requirement then reads as

$$E = 0.08\left[\tfrac{1}{2}G + N + \sum_j \max\{(|x_j| - 0.2G), 0\}\right] \qquad (3)$$

Paradoxically, while the surface $C$ = const. can be shown to be convex in any dimension, the extra term, which was designed precisely to prevent excessive concentration of the portfolio, can make the level surfaces of $E$ concave in some regions of weight space, provided the dimension is higher than a threshold. This threshold dimension depends on the factor in front of $G$ in the extra term; for the value 0.2 in the actual regulation, convexity is lost at and above 4 dimensions. High dimensional surfaces are not easy to display, therefore we suggest that the reader convince herself about the concavity of the surface by considering some simple portfolios like
$a = \{0.95; -1.05; 1; 0.1\}$, $b = \{1; -1; 1; 0\}$, and $c = \{1.05; -0.95; 1; -0.1\}$ for which the capital requirement works out to be $E(a) = E(c) = 0.2952$, and $E(b) = 0.296$, respectively. Since $b$ is the mean of $a$ and $c$, these values evidently violate convexity, albeit fairly weakly. In higher dimensions the effect is much stronger than in the above example, which we have chosen only in order to show that a concave region can appear already in 4 dimensions.

## 2.4. The general risk of bonds

The rules described in Annex I, §§15 to 30 are trying to capture the risk of the bond portfolio due to a shift and distortion of the interest rate curve. There are two methods, one based on maturity (§§15 to 23), the other on duration (§§24 to 30). Both methods are fairly complicated; here we give details only for the simpler, the one based on duration, and content ourselves with simply announcing the result for the other.

The calculation of the capital charge starts by sorting the positions into three zones within and between which a certain amount of capital reduction is allowed according to a recursively defined matching procedure.

The algorithm is best explained on the scheme shown in Table I.

| Zone 1 | $L_1$ | $S_1$ | $M_1$ | $U_1$ | $M_{1,2}$ | $U_1'$ |        | $U_1''$ |           |
|--------|-------|-------|-------|-------|-----------|--------|--------|----------|-----------|
| Zone 2 | $L_2$ | $S_2$ | $M_2$ | $U_2$ |           | $U_2'$ | $M_{2,3}$ | $U_2''$ | $M_{1,3}$ |
| Zone 3 | $L_3$ | $S_3$ | $M_3$ | $U_3$ |           |        |        | $U_3'$   | $U_3''$   |

Table I: An aid to the algorithm for the calculation of the capital requirement of the bond portfolio by the duration-based method.

As a first step, the long ( $L_j$ ) and short ( $S_j$ ) positions are calculated for zone $j$ from the individual positions $x_i^{(j)}$ of item $i$ in zone $j$ as

$$L_j = \sum_i \frac{1}{2}\left(sign(x_i^{(j)}) + 1\right) \cdot x_i^{(j)}$$

and

$$S_j = \sum_i \frac{1}{2}\left(sign(x_i^{(j)}) - 1\right) \cdot x_i^{(j)} \quad (>0) ,$$

respectively. Next the matched position $M_j$ is worked out for zone $j$ as

$$M_j = \min\{S_j, L_j\} = \frac{1}{2}\left(L_j + S_j - |L_j - S_j|\right) = \frac{1}{2}\sum_i |x_i^{(j)}| - \frac{1}{2}\left|\sum_i x_i^{(j)}\right|$$

The unmatched positions $U_j$ will be given by

$$U_j = \sum_i x_i^{(j)}$$

and can be of either sign. Now matching is repeated between zones 1,2 , resp. 2,3 , resp. 1,3:

The matched position between zones 1,2 will be

$$M_{1,2} = \min\{|U_1|,|U_2|\} = \frac{1}{2}(|U_1|+|U_2|) - \frac{1}{2}|U_1+U_2|,$$

with the corresponding unmatched positions

$$U_1' = signU_1(|U_1| - M_{1,2}),$$

$$U_2' = signU_2(|U_2| - M_{1,2}).$$

Similarly, the matched position between zones 2,3 is

$$M_{2,3} = \frac{1}{2}(|U_2'|+|U_3|) - \frac{1}{2}|U_2'+U_3|$$

and the unmatched ones are

$$U_2'' = signU_2'(|U_2'| - M_{2,3}),$$

$$U_3' = signU_3(|U_3| - M_{2,3}).$$

Finally, the matching between zones 1,3 yields

$$M_{1,3} = \frac{1}{2}(|U_1'|+|U_3'|) - \frac{1}{2}|U_1'+U_3'|$$

$$U_1'' = signU_1'(|U_1'| - M_{1,3})$$

$$U_3'' = signU_3'(|U_3'| - M_{1,3}).$$

Finally, the capital requirement in the duration-based method is given by

$$B_d = \lambda \sum_{j=1}^{3} M_j + \lambda_{1,2}(M_{1,2} + M_{2,3}) + \lambda_{1,3}M_{1,3} + |U_1''| + |U_2''| + |U_3''|, \qquad (4)$$

where for the weights we have $\lambda = 0.02$, $\lambda_{1,2} = 0.4$, and $\lambda_{1,3} = 1.5$.

In the maturity-based method we have, in addition to the zones, also different bands within the zones. With $M_j^i$ the matching position within band $i$ in zone $j$ and $\lambda_0 = 0.1$, $\lambda_1 = 0.4$, $\lambda_2 = \lambda_3 = 0.3$, $\lambda_{1,2} = 0.4$, and $\lambda_{1,3} = 1.5$, we find that the capital requirement is

$$B_m = \lambda_0 \sum_{i,j} M_j^i + \lambda_1 M_1 + \lambda_2 M_2 + \lambda_3 M_3 + \lambda_{1,2}(M_{1,2} + M_{2,3}) + \lambda_{1,3} M_{1,3} + |U_1''| + |U_2''| + |U_3''|$$

(5)

An examination of the $B = $ const. surfaces reveals that they can be deeply concave in both cases. For example, for a portfolio consisting of just three bonds belonging to three different zones the iso-risk surface in the duration-based method looks like that shown in Fig. 3. The geometry of the iso-risk surfaces belonging to the general risk of bonds is very rich. A gallery of risk surfaces corresponding to portfolios with various distribution of bonds over the zones can be found at www.colbud.hu/kondor/risksurfaces.

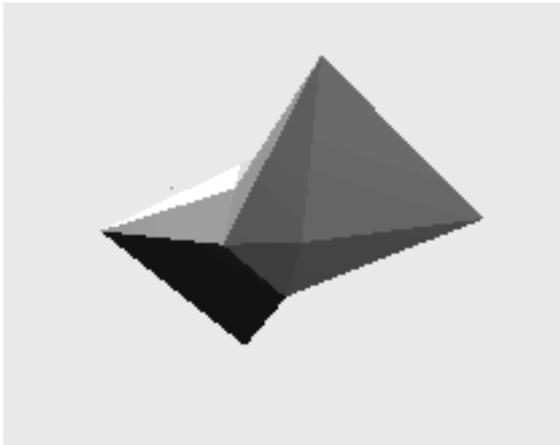

Fig.3: The level surface of the general risk of a bond portfolio consisting of three bonds belonging to three different zones in the duration-based method.

The inconsistency is due to the particular choice of the weights, specifically to the disproportionately large weight associated with the matching between zones 1 and 3.

The large differences between the weights are the source of a further inconsistency: whenever a large enough component of an otherwise fixed composition portfolio crosses a zone boundary, there is a huge jump in the capital charge.

We have performed a study of the behaviour of the $B = $ const. surfaces, regarding the weights as essentially free parameters. After an extremely tedious analysis that we do not have sufficient space to reproduce here, we found the surprisingly simple result that there always exist bond portfolios for which the iso-risk surfaces of the general risk will be concave whenever the following inequalities break down:

$$\lambda_{1,3} \leq 2\lambda_{1,2} - \lambda \quad,$$

(6)

for the duration-based, and

$$\lambda_{1,3} \leq 2\lambda_{1,2} - \lambda_2 \tag{7}$$

for the maturity-based method. Note that in the latter case, of the weights belonging to a given zone, only the one associated to the middle zone, $\lambda_2$, enters the condition (7). It is evident that for the weights defined in the regulation these inequalities are violated both in the maturity-based and in the duration-based method, which explains the lack of convexity in both cases. The minimal change of the regulation that would restore convexity would be a modification of the values of the weights. Keeping everything else constant the weights associated with the matching position between zones 1 and 3 should be reduced to below 78% and 50% in the duration-based and in the maturity-based method, respectively, in order to avoid the appearance of concave regions on the iso-risk surfaces.

The proof of the inequalities (6), (7) will be published elsewhere.

## 3. Conclusion

We have studied the risk measures implied by current market risk capital regulation. We found that the surfaces of constant capital charge (constant risk) are polyhedra. This is due to the fact that the regulator obviously wanted to construct these risk measures from relatively simple blocks, like the gross and net positions. In the case of the specific risk of bonds and that of the FX portfolio, the risk measures implied by the regulation are convex, so they at least satisfy an obvious necessary condition. In the case of the equity portfolio and that of the general interest rate risk, we found violations of convexity. This inconsistency could easily be remedied by a modification of the weight parameters appearing in the formulae.

It is not entirely clear how important these inconsistencies are in actual practice. It is conceivable that in the case of a large portfolio the concave regions, and maybe also the wild jumps in the capital charge of the bond portfolio, are washed away. It is also possible that compared with the potential damages of using VaR for a risk measure, the consequences of the lack of convexity in the risk measures described here are of minor importance. Nevertheless, our findings reinforce the impression that international capital regulation notoriously lacks a solid theoretical foundation, and *that in itself* constitutes a major element of systemic risk.

**Acknowledgement**
Thanks are due to P. Ittzes for technical assistance, and J. Steger for generating the figures. One of us (I.K.) has been partially supported by the Hungarian Research Foundation OTKA, grant No. T-034835.